\documentclass[aps,prl,singlecolumn,groupedaddress,showpacs,amsmath]{revtex4}

\usepackage{epsfig}

\makeatletter
\def\@dotsep{4.5}
\makeatother
\newtheorem{lemma}{Lemma}
\newtheorem{theorem}{Theorem}

\newcommand{\eql}[1]{\label{eq:#1}}
\newcommand{\eqr}[1]{(\ref{eq:#1})}

\begin{document}

%\preprint{}

\title{Theorem on subwavelength imaging with arrays of discrete sources}


\author{S. I. Maslovski}
\email[]{stas@co.it.pt}
%\thanks{}
\affiliation{Departamento de Engenharia Electrot\'{e}cnica\\
Instituto de Telecomunica\c{c}\~{o}es, Universidade de Coimbra\\
P\'{o}lo II, 3030-290 Coimbra, Portugal}
%\altaffiliation{On leave from Radiophysics Dept.\\
%St. Petersburg State Polytechnical University\\
%Politekhnicheskaya 29, 195251, St.~Petersburg, Russia}


\date{\today}

\begin{abstract}
  A theorem on subwavelength imaging with arrays of discrete
  sources is formulated. This theorem is analogous to the
  Kotelnikov (also named Nyquist-Shannon) sampling theorem as it represents
  the field at an arbitrary point of space in terms of the same field
  taken at discrete points and imposes similar limitations on the accuracy of the image.
  A physical realization of an imaging system operating exactly on the resolution limit
  enforced by the theorem is outlined. 
\end{abstract}

%\keywords{}

\maketitle


In last decade, certain interest has appeared to electromagnetic systems
that, under specific conditions, are able to produce electromagnetic
fields localized in areas of characteristic dimensions much less than
the wavelength of the used radiation. Different physical mechanisms
can be applied to achieve this, ranging from slabs of Veselago media
and systems analogous to them
\cite{Pendry_lens_PRL_2000,Grbic_lens_PRL_2004,Alitalo_lens_JAP_2006,Alitalo_explens_JAP_2006} to
impedance grids or arrays
\cite{Maslovski_polariton_JAP_2004,Freire_lens_APL_2005}
or even nonlinear systems \cite{Maslovski_conjugation_JAP_2003,Pendry_reversal_Science_2008}.

The mentioned systems deal with quickly decaying near fields of a
source, i.e., with the evanescent modes of the source field.
Merlin {\it et al.} \cite{Merlin_focusing_Science_2007,Grbic_focusing_Science_2008}
proposed to transform a part of the
propagating spectrum to the evanescent spectrum with the help of a
planar membrane that modulates the incident field amplitude and
phase. The membrane can be designed in such a way that the evanescent
waves produced by it form a ``beam'' of sub-wavelength dimensions at a
certain distance from the array. Merlin named this phenomenon {\em
  radiationless interference}. Physically, interference here means
superimposing the fields of the secondary sources associated with the
illuminated membrane. In \cite{Merlin_focusing_Science_2007} a special
distribution of these sources was considered with which the
subwavelength field localization can be achieved.

Another interference-based method of subwavelength focusing was proposed in
\cite{Eleftheriades_holography_I3EMWCL_2008,Markley_focusing_PRL_2008,Markley_focusing_I3EMWCL_2009}.
In this approach the incident field penetrates trough a number of slots in a metal screen.
The beam patterns produced by displaced slot elements form a set of basis functions in which
the image-plane field distribution can be expanded and the necessary beam formation can be achieved.

In this letter we will study limitations on subwavelength imaging imposed by discrete nature
of the considered sources. We will establish a theorem in the space domain that can be seen
as an analogy of the well-known Kotelnikov (Nyquist-Shannon) sampling theorem in the time
domain~\cite{Kotelnikov_sampling_RKKA_1933,Shannon_noise_PIRE_1949}. We will also outline a possible
realization of an imaging system operating exactly on the resolution limit enforced by this theorem.

Let us start with considering an infinite planar array of high-frequency line currents in free space.
We introduce a Cartesian coordinate system with the $x$-axis oriented
along the currents and the $z$-axis perpendicular to the array
plane. In these coordinates the currents are at points $z = 0$, $y =
nl$,
where $l$ is the period of the array and $n\in {\cal Z}$. The
complex amplitudes of the currents are yet unknown, let us denote them
$I_n$. Then the total electric field at the point $(0, y, z)$ produced
by this array of currents reads:
\begin{equation}
E_x(y, z) = -{k\eta\over 4}\sum\limits_{n = -\infty}^{+\infty}{I_n H_0^{(2)}(k\sqrt{(nl - y)^2 + z^2})},
\eql{field}
\end{equation} 
where $k = 2\pi/\lambda$, $\eta = \sqrt{\mu_0/\varepsilon_0}$; the time dependence is of form $\exp(+j\omega t)$.

\begin{figure}[tb]
\centering
\epsfig{file=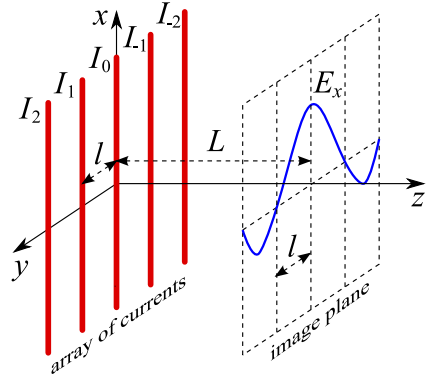,width=5.5cm}
\caption{(color online) An array of line currents at §z=0§ with the image plane at $z=L$. We demand the electric field $E_x$ vanishing
at discrete points $z=L$, $y=ml$, $m \neq 0$.}
\label{array}
\end{figure}

Let us impose the following condition on the electric field at the plane $z=L$ (image plane):
\begin{equation}
\begin{array}{lll}
E_x(0, L) & = & E_0,\\
E_x(ml, L) & = & 0, \quad m\neq 0. 
\end{array}
\eql{condition}
\end{equation}
The respective array and the image plane are shown in Fig.~\ref{array}.
Physically, Eqs.~\eqr{condition} mean that we would like to concentrate the field at
the image plane around point $y=0$ and at the same time have it
completely vanishing at discrete points $y = ml$, $m\neq 0$. Next, we
will look for such a distribution of currents that satisfies
\eqr{condition}.

Considering a Fourier series with the discrete currents $I_n$ as its coefficients one can write
\begin{equation}
I_n = {1\over 2\pi}\int\limits_{-\pi}^{\pi}{\cal J}(q)e^{-jnq}dq,
\eql{fourier}
\end{equation}
where ${\cal J}(q)$ is yet unknown function. Substituting
\eqr{fourier} into \eqr{field} we obtain
\begin{multline}
E_x(ml, L) = -{k\eta\over 8\pi} \sum\limits_{n=-\infty}^{+\infty}
H_0^{(2)}(k\sqrt{(n-m)^2l^2+L^2})\int\limits_{-\pi}^{\pi}{\cal J}(q)e^{-jnq}dq\\
=-{k\eta\over 8\pi}\int\limits_{-\pi}^{\pi}{\cal J}(q)e^{-jmq}
\sum\limits_{n=-\infty}^{+\infty}H_0^{(2)}(k\sqrt{n^2l^2+L^2})e^{-jnq}\,dq.
\eql{derivation}
\end{multline}

Next, using the Poisson summation formula
\begin{equation}
\sum\limits_{n=-\infty}^{+\infty}H_0^{(2)}(k\sqrt{n^2l^2+L^2})e^{-jnq}\,dq
= 2j\sum\limits_{m=-\infty}^{+\infty}{e^{-{L\over l}\sqrt{(2\pi m - q)^2-k^2l^2}}\over\sqrt{(2\pi m - q)^2-k^2l^2}}
\equiv {\cal K}(q),
\eql{poisson}
\end{equation}
where ${\cal K}(q)$ is a quickly converging series; the branch of the
square root in \eqr{poisson} is such that $\sqrt{-1} = +j$. It is seen
that ${\cal K}(-q) = {\cal K}(q)$ and ${\cal K}(q + 2\pi n) = {\cal
  K}(q)$.

From \eqr{derivation} and \eqr{poisson} we get
\begin{equation}
E_x(ml, L) = -{k\eta\over 8\pi}\int\limits_{-\pi}^{\pi}{\cal J}(q){\cal K}(q)e^{-jmq}\,dq.
\eql{final}
\end{equation}
From \eqr{final} and the condition \eqr{condition} it immediately
follows that
\begin{equation}
{\cal J}(q) = -{4E_0\over k\eta}{\cal K}(q)^{-1}.
\eql{solution}
\end{equation}
Finally, sustituting \eqr{solution} into \eqr{fourier} we get the complex amplitudes of the currents:
\begin{equation}
I_n = -{4E_0\over k\eta}\Psi(n), \quad \Psi(n) =
{1\over \pi}\int\limits_{0}^{\pi}{\cos nq \over {\cal K}(q)}\,dq,
\eql{waveletdef}
\end{equation}
where we have introduced the function $\Psi(n)$ that we will call
{\em elementary wavelet}. In terms of it, condition \eqr{condition} can be
rewritten as
\begin{equation}
\sum\limits_{n=-\infty}^{+\infty}\Psi(n) H_0^{(2)}(k\sqrt{(n-m)^2l^2 + L^2}) =
\left\{
\begin{array}{lcl}
1, \quad m &=& 0,\\
0, \quad m &\neq & 0.
\end{array}
\right.
\eql{wavelets}
\end{equation}
Now we can formulate a couple of lemmas.

\begin{lemma}
  For any given electric field distribution $E_x(y_n)$ defined at
  discrete points $y_n = nl$ in the image plane there exists a
  distribution of line currents defined at the points $y_n = nl$ in
  the source plane that produces exactly the given electric field
  distribution. This distribution of currents is
\begin{equation}
I_n = -{4\over k\eta}\sum\limits_{m = -\infty}^{\infty}E_x(ml)\Psi(n - m).
\eql{expansion1}
\end{equation}
\label{lemma1}
\end{lemma}

This lemma is obvious: it directly follows from the superposition
principle and the solution of the problem considered above.

\begin{lemma}
  $\Psi(n - m)$ form a complete system of linearly independent
  functions on an infinite set of discrete points so that any given
  distribution of line currents $I_n$ can be uniquely expanded into
  wavelet series
\begin{equation}
I_n = \sum\limits_{m = -\infty}^{\infty} C_m \Psi(n - m).
\eql{expansion}
\end{equation}
\label{lemma2}
\end{lemma}

The proof of this lemma is given in Appendix.

Now from the above lemmas we can conclude that the distribution of the
field produced by a planar periodic array of discrete sources is
completely determined by the values of the same field taken at
corresponding discrete points in any plane parallel to the array
plane. Indeed, knowing the discrete values of the field one can
reconstruct the currents in the source plane with the help of Lemma
\ref{lemma1} and then calculate the field at any point of space
using~\eqr{field}. Lemma \ref{lemma2} makes sure that the source
distribution found in this way is unique.

Having said that we arrive at the following theorem which is in a
sense analogous to the Kotelnikov sampling theorem.

\begin{theorem}
  The electric field of the considered planar periodic array of line
  currents is completely determined by the values of the same field
  taken at certain discrete points in any plane parallel to the array
  plane and can be represented by the following formula:
\begin{equation}
E_x(y, z) = \sum\limits_{m = -\infty}^{+\infty}E_x(ml,L)
\sum\limits_{n = -\infty}^{+\infty}\Psi(n - m)H_0^{(2)}(k\sqrt{(nl - y)^2 + z^2}).
\eql{theorem}
\end{equation}
\end{theorem}
As said above, this formula is the result of substitution of
Eq.~\eqr{expansion1} into Eq.~\eqr{field}.

From this theorem we see that it is impossible to produce {\it arbitrary}
near-field distributions with an array of currents of a given period.
The set of realizable distributions is determined by \eqr{theorem} and
is, basically, a set of linear combinations of the fields created by
the wavelets of currents \eqr{expansion}. Thus, the structure
period sets a limit on the amount of details that one can expect to
observe in a near-field image. Although this fact is rather physically
intuitive and has been mentioned in the literature before, the above theorem
makes it explicit in mathematical terms.

Let us now discuss some properties of the introduced wavelet
functions. The elementary wavelet is given by~\eqr{waveletdef}. When
the period of the array is small compared to the wavelength and the
parameter $L/l$ is large enough, the series ${\cal K}(q)$ converges
very rapidly, so that only a few first terms of the series should be
taken into account. For instance, for $kl \le \pi/2$ and $L/l \ge 3$ it is
enough to take only 3 terms of the series.

\begin{figure}[tb]
\centering
\epsfig{file=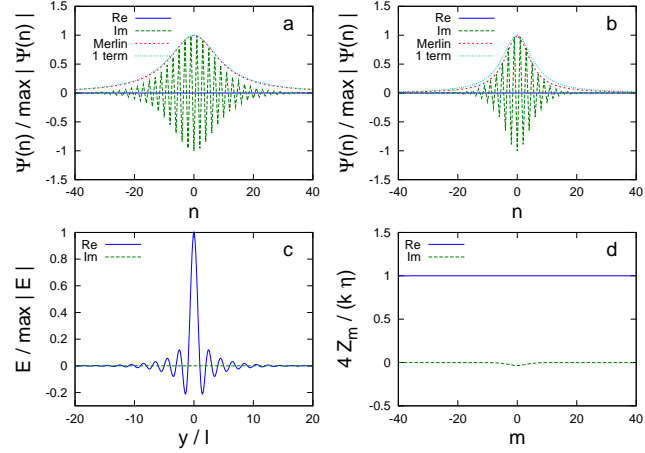, width=8.6cm}
\caption{(color online) (a, b) The normalized elementary wavelets $\Psi(n)$ as functions of $n$.\
The parameters are: a) $L/l = 10$, $kl = 0.1\pi$; b) $L/l = 5$, $kl = 0.45\pi$.
(c) The normalized electric field $E_x$ in the image plane as a function of $y/l$.\ 
The parameters are as in Fig.~\ref{waveletfig}(b).
(d) The second term of the total wire impedance $Z_m^{\rm tot}$ normalized to $k\eta/4$ as a function of $m$.}
\label{waveletfig}
\end{figure}

Example plots of the normalized elementary wavelets for two cases with different values of the parameters
are given in Fig.~\ref{waveletfig}(a,~b). One can see that the normalized $\Psi(n)$ is
practically purely imaginary, as the real part of it approaches zero for $kl \le \pi/2$ and $L/l \ge 3$.
On the plot there are also shown the distribution
of current magnitudes of the source introduced by Merlin
\cite{Merlin_focusing_Science_2007}: $|I_n|\propto 1/(1+n^2l^2/L^2)$ and the distribution
that we can obtain from our formulas when only a single term of the
series \eqr{poisson} with $m = 0$ is taken into account. We conclude that when $L/l \gg 1$ the source proposed
by Merlin can be seen as a one-term approximation of our
formulas. This approximation is valid only for small $n$ because
$\Psi(n)$ decays quicker (exponentially) when
$n\rightarrow\infty$.
 
The field distribution in the image plane is depicted in Fig.~\ref{waveletfig}(c).
This function reminds of $\mbox{sinc}(y/l)$, however for large
arguments it decays quicker (exponentially). One can see that the
width of the near-field peak at zero level is $2l$, as expected.

Next we would like to consider how the introduced wavelets (which are
certain distributions of line currents) can be physically realized. It
is natural to represent a two-dimensional array of line currents with a
grid of thin wires illuminated by an incident plane wave. As we would like to excite currents with
alternating phases on the neighboring wires (see Fig.~\ref{waveletfig}(a,b))
it is natural to load the wires with alternating reactive loads, so
that the effective wire impedances and wire currents will change from
wire to wire in the necessary way. Contrary to the scheme proposed
in \cite{Grbic_focusing_Science_2008} we would like to separate the
creation of the phase profile of the currents from the amplitude
profile. Indeed, the characteristic scale of the amplitude profile $L$ can
be much larger than the phase change scale $l$. In the example given in
Fig.~\ref{waveletfig}(a) these scales differ by 10 times, so that the phase scale can
be subwavelength while the amplitude scale is not. Therefore, it
is more practical to realize the amplitude profile with the
conventional optical means by shaping the incident beam with
diaphragms and/or lenses, while the phase profile can be realized with
wire loading.

As a proof of concept let us calculate the interaction field and the
local field in a wire array with currents distributed proportionally to $\Psi(n)$:
$I_n = I_0[\Psi(n)/\Psi(0)]$ for the case depicted in Fig.~\ref{waveletfig}(b).
The interaction field at the $m$-th wire is
\begin{equation}
E_x^{\rm int}(ml,0) = -{k\eta I_0\over 4\Psi(0)}\sum
\limits_{n=-\infty\atop n\neq m}^{+\infty}
\Psi(n)H_0^{(2)}(|n-m|kl).
\end{equation}
Given the incident field $E^{\rm inc}_x(y,z)$ the local field acting
on the $m$-th wire can be expressed as
$
E^{\rm loc}_m = E^{\rm inc}_x(ml,0) + E_x^{\rm int}(ml,0).
$
As we discussed above, we realize the necessary amplitude profile of the
currents by shaping the incident field accordingly, so that
$
E^{\rm inc}_x(ml,0) = E^{\rm inc}_0 {|\Psi(m)|/\Psi(0)}.
$
Therefore, the total effective impedance (including the
self-impedance and loading) per unit length of the $m$-th wire has to be
\begin{equation}
Z^{\rm tot}_m = {E^{\rm loc}_m\over I_m} = {E^{\rm inc}_0\over I_0} {|\Psi(m)|\over\Psi(m)}\\
-{k\eta \over 4\Psi(m)}\sum
\limits_{n=-\infty\atop n\neq m}^{+\infty}
\Psi(n)H_0^{(2)}(|n-m|kl).
\eql{ztot}
\end{equation}
The phase of the first term in \eqr{ztot} is alternating from wire to
wire while its magnitude remains constant. Moreover, if $I_0$
is chosen \footnote{One can always provide that with appropriate loading.
That is exactly what we do in this section: we search for such a loading
that will give us the necessary distribution of currents.} to be in phase with $E^{\rm inc}_0$ then this impedance term
is practically purely reactive (imaginary). The dependence of the
second term on wire index is plotted in Fig.~\ref{waveletfig}(d).
One can see that the real part of this term is constant and equals
the radiation resistance of a single wire. Indeed, for the self-impedance of a thin wire of radius $r$ we have \cite{Tretyakov_modelling_2003}
$4Z_{\rm self}/(k\eta) = H_0^{(2)}(kr) \approx 1 + j(2/\pi)\log[2/(\gamma kr)]$, where $\log(\gamma) \approx 0.5772$.

Another important property of it seen from Fig.~\ref{waveletfig}(b) is that the imaginary part is also
practically constant with~$m$~\footnote{We have numerically checked this for various combinations of parameters $kl \le \pi/2$ and $L/l \ge 3$.}.
Moreover, for the chosen parameters it happens to be very close to zero.
This means that in a real array of wires we can neglect the small difference in this part of the impedance and
load {\it all} wires by loads of only two kinds, to account for
alternating reactance of the constant magnitude in the first term of
$Z_m^{\rm tot}$. When compared to \cite{Grbic_focusing_Science_2008}
this greatly simplifies the structure. Moreover, our loaded wire grid
is periodic with the period $2l$, therefore if the beam illuminating
the array moves along $y$-axis by an integral number of periods,
the subwavelength spot produced by the array will
also move by the same number of periods. This important property is
missing in the realizations proposed
in~\cite{Grbic_focusing_Science_2008,Eleftheriades_holography_I3EMWCL_2008,Markley_focusing_PRL_2008,Markley_focusing_I3EMWCL_2009}.

To conclude, in this letter we have formulated a theorem on
subwavelength imaging which is in a sense analogous to the Kotelnikov
sampling theorem. In the process of derivation of this theorem we
solved a problem about a distribution of currents in a two-dimensional
array that produces a given subwavelegth near-field spot.
The found distribution of line currents represents an elementary source wavelet.
It has been shown that the source proposed in
\cite{Merlin_focusing_Science_2007,Grbic_focusing_Science_2008}
can be understood as a one-term approximation of the elementary wavelet derived in this letter.
We have also shown that if the necessary amplitude profile of the currents is realized by shaping
the incident field by conventional optical means the remaining phase
profile can be produced with a grid of thin wires loaded by reactive loads of only two kinds.
When compared to structures considered by other authors such a periodically loaded wire grid
appears to be much simpler to realize.

\section*{Appendix}

We start with the proof of linear independence. Let us assume that
there exists a non-trivial linear combination of wavelets such that
\begin{equation}
\sum_{m = -\infty}^{\infty} \alpha_m\Psi(n - m) = 0, \quad \forall\,n\in {\cal Z}.
\eql{combo}
\end{equation}
This linear combination defines a source distribution with all
currents equal to zero: $I_n \equiv 0$, therefore, the field exited by
these vanishing currents must also vanish everywhere.

From the other hand, \eqr{combo} is a non-trivial linear combination,
therefore there exists $\alpha_p \neq 0$. But then from the definition
of $\Psi(n)$ it immediately follows that the electric field in the
image plane at the point $y_p = pl$ is $E_x(y_p) = -(k\eta/4)\alpha_p
\neq 0$.
We have arrived at a contradiction, therefore $\Psi(n - m)$
are linearly independent.

Completeness of the system of functions $\Psi(n - m)$ can be seen from
the following. Consider the linear combination of wavelets
\eqr{expansion} with coefficients given by $C_m = H_0^{(2)}(k\sqrt{(p
  - m)^2l^2+L^2})$.
Then, denoting $s = n - m$ and using~\eqr{wavelets} we get for
the currents
\begin{equation}
I_n =\sum_{s=-\infty}^{+\infty}H_0^{(2)}(k\sqrt{(s - (n - p))^2l^2+L^2})\Psi(s)
=\left\{
\begin{array}{lcl}
1, \quad n &=& p,\\
0, \quad n &\neq & p.
\end{array}
\right.
\end{equation}
Since $p$ is arbitrary we see that the current localized at any point
$y_p = pl$ can be expressed in terms of the wavelets, therefore any
distribution of such currents can be expressed also.

%\bibliography{references}

%\newpage
%\listoffigures

\end{document}